%% file: main.tex
\DeclareRobustCommand{\ion}[2]{%
\relax\ifmmode
\ifx\testbx\f@series
{\mathbf{#1\,\mathsc{#2}}}\else
{\mathrm{#1\,\mathsc{#2}}}\fi
\else\textup{#1\,{\mdseries\textsc{#2}}}%
\fi}

\documentclass{webofc}
\usepackage[varg]{txfonts}   
\usepackage{graphicx}
\usepackage{subfig}
\usepackage{xcolor}
\usepackage[colorlinks=true,allcolors=blue]{hyperref}
\usepackage{url}
\usepackage{caption}
\begin{document}
\title{Exploring the interstellar medium of NGC~891 at millimeter wavelengths using the NIKA2 camera}
%
%

\author{%
  \lastname{S.~Katsioli}\inst{\ref{Athens_obs},\ref{Athens_univ}}\fnsep\thanks{e-mail: s.katsioli@noa.gr}
  \and  R.~Adam \inst{\ref{OCA}}
  \and  P.~Ade \inst{\ref{Cardiff}}
  \and  H.~Ajeddig \inst{\ref{CEA}}
  \and  P.~Andr\'e \inst{\ref{CEA}}
  \and  E.~Artis \inst{\ref{LPSC},\ref{Garching}}
  \and  H.~Aussel \inst{\ref{CEA}}
  \and  M.~Baes  \inst{\ref{Ghent}}
  \and  A.~Beelen \inst{\ref{LAM}}
  \and  A.~Beno\^it \inst{\ref{Neel}}
  \and  S.~Berta \inst{\ref{IRAMF}}
  \and  L.~Bing \inst{\ref{LAM}}
  \and  O.~Bourrion \inst{\ref{LPSC}}
  \and  M.~Calvo \inst{\ref{Neel}}
  \and  A.~Catalano \inst{\ref{LPSC}}
  \and  C.~J.~R.~Clark \inst{\ref{STSI}}
  \and  I.~De~Looze  \inst{\ref{Ghent}, \ref{UCL}}
  \and  M.~De~Petris \inst{\ref{Roma}}
  \and  F.-X.~D\'esert \inst{\ref{IPAG}}
  \and  S.~Doyle \inst{\ref{Cardiff}}
  \and  E.~F.~C.~Driessen \inst{\ref{IRAMF}}
  \and  G.~Ejlali \inst{\ref{Tehran}}
  \and  M.~Galametz  \inst{\ref{CEA}}
  \and  F.~Galliano  \inst{\ref{CEA}}
  \and  A.~Gomez \inst{\ref{CAB}} 
  \and  J.~Goupy \inst{\ref{Neel}}
  \and  C.~Hanser \inst{\ref{LPSC}}
  \and  A.~Hughes  \inst{\ref{Toulouse}}
  \and  F.~K\'eruzor\'e \inst{\ref{Argonne}}
  \and  C.~Kramer \inst{\ref{IRAMF}}
  \and  A.~P.~Jones  \inst{\ref{IAS}}
  \and  B.~Ladjelate \inst{\ref{IRAME}} 
  \and  G.~Lagache \inst{\ref{LAM}}
  \and  S.~Leclercq \inst{\ref{IRAMF}}
  \and  J.-F.~Lestrade \inst{\ref{LERMA}}
  \and  J.~F.~Mac\'ias-P\'erez \inst{\ref{LPSC}}
  \and  S.~C.~Madden \inst{\ref{CEA}}
  \and  A.~Maury \inst{\ref{CEA}}
  \and  P.~Mauskopf \inst{\ref{Cardiff},\ref{Arizona}}
  \and  F.~Mayet \inst{\ref{LPSC}}
  \and  A.~Monfardini \inst{\ref{Neel}}
  \and  A.~Moyer-Anin \inst{\ref{LPSC}}
  \and  M.~Mu\~noz-Echeverr\'ia \inst{\ref{LPSC}}
  \and  A.~Nersesian  \inst{\ref{Ghent}, \ref{Athens_obs}}
  \and  L.~Pantoni  \inst{\ref{CEA}, \ref{IAS}}
  \and  D.~Paradis  \inst{\ref{Toulouse}}
  \and  L.~Perotto \inst{\ref{LPSC}}
  \and  G.~Pisano \inst{\ref{Roma}}
  \and  N.~Ponthieu \inst{\ref{IPAG}}
  \and  V.~Rev\'eret \inst{\ref{CEA}}
  \and  A.~J.~Rigby \inst{\ref{Leeds}}
  \and  A.~Ritacco \inst{\ref{INAF}, \ref{ENS}}
  \and  C.~Romero \inst{\ref{Pennsylvanie}}
  \and  H.~Roussel \inst{\ref{IAP}}
  \and  F.~Ruppin \inst{\ref{IP2I}}
  \and  K.~Schuster \inst{\ref{IRAMF}}
  \and  A.~Sievers \inst{\ref{IRAME}}
  \and  M.~W.~L.~Smith  \inst{\ref{Cardiff}}
  \and  J.~Tedros \inst{\ref{IRAME}}
  \and  F.~Tabatabaei  \inst{\ref{Tehran}}
  \and  C.~Tucker \inst{\ref{Cardiff}}
  \and  E.~M.~Xilouris  \inst{\ref{Athens_obs}}
  \and  N.~Ysard  \inst{\ref{IAS}}
  \and  R.~Zylka \inst{\ref{IRAMF}}
}
\institute{
  National Observatory of Athens, IAASARS, GR-15236, Athens, Greece
  \label{Athens_obs}
  \and
  Faculty of Physics, University of Athens, GR-15784 Zografos, Athens, Greece
  \label{Athens_univ}
  \and
  Universit\'e C\^ote d'Azur, Observatoire de la C\^ote d'Azur, CNRS, Laboratoire Lagrange, France 
  \label{OCA}
  \and
  School of Physics and Astronomy, Cardiff University, CF24 3AA, UK
  \label{Cardiff}
  \and
  Universit\'e Paris-Saclay, Université Paris Cité, CEA, CNRS, AIM, 91191, Gif-sur-Yvette, France
  \label{CEA}
  \and
  Universit\'e Grenoble Alpes, CNRS, Grenoble INP, LPSC-IN2P3, 38000 Grenoble, France
  \label{LPSC}
  \and	
  Max Planck Institute for Extraterrestrial Physics, 85748 Garching, Germany
  \label{Garching}
  \and
  Sterrenkundig Observatorium Universiteit Gent, Krijgslaan 281 S9, B-9000 Gent, Belgium
  \label{Ghent}
  \and
  Aix Marseille Univ, CNRS, CNES, LAM, Marseille, France
  \label{LAM}
  \and
  Universit\'e Grenoble Alpes, CNRS, Institut N\'eel, France
  \label{Neel}
  \and
  Institut de RadioAstronomie Millim\'etrique (IRAM), Grenoble, France
  \label{IRAMF}
  \and
  Space Telescope Science Institute, 3700 San Martin Drive, Baltimore, MD 21218, USA
  \label{STSI}
  \and
  Department of Physics and Astronomy, UCL, Gower Street, London WC1E 6BT, UK
  \label{UCL}
  \and 
  Dipartimento di Fisica, Sapienza Universit\`a di Roma, I-00185 Roma, Italy
  \label{Roma}
  \and
  Univ. Grenoble Alpes, CNRS, IPAG, 38000 Grenoble, France
  \label{IPAG}
  \and
  Institute for Research in Fundamental Sciences (IPM), Larak Garden, 19395-5531 Tehran, Iran
  \label{Tehran}
  \and
  Centro de Astrobiolog\'ia (CSIC-INTA), Torrej\'on de Ardoz, 28850 Madrid, Spain
  \label{CAB}
  \and
  IRAP, Université de Toulouse, CNRS, UPS, IRAP, BP 44346, 31028 Toulouse Cedex 4, France
  \label{Toulouse}
  \and
  High Energy Physics Division, Argonne National Laboratory, Lemont, IL 60439, USA
  \label{Argonne}
  \and
  Universit\'e Paris-Saclay, CNRS, Institut d'astrophysique spatiale, 91405, Orsay, France
  \label{IAS}
  \and  
  Instituto de Radioastronom\'ia Milim\'etrica (IRAM), Granada, Spain
  \label{IRAME}
  \and
  LERMA, Observatoire de Paris, PSL, CNRS, Sorbonne Univ., UPMC, 75014 Paris, France  
  \label{LERMA}
  \and
  School of Earth \& Space and Department of Physics, Arizona State University, AZ 85287, USA
  \label{Arizona}
  \and
  School of Physics and Astronomy, University of Leeds, Leeds LS2 9JT, UK
  \label{Leeds}
  \and
  INAF-Osservatorio Astronomico di Cagliari, 09047 Selargius, Italy
  \label{INAF}
  \and
  LPENS, ENS, PSL Research Univ., CNRS, Sorbonne Univ., Universit\'e de Paris, 75005 Paris, France 
  \label{ENS}
  \and    
  Department of Physics and Astronomy, University of Pennsylvania, PA 19104, USA
  \label{Pennsylvanie}
  \and
  Institut d'Astrophysique de Paris, CNRS (UMR7095), 75014 Paris, France
  \label{IAP}
  \and
  University of Lyon, UCB Lyon 1, CNRS/IN2P3, IP2I, 69622 Villeurbanne, France
  \label{IP2I}
}

\abstract{%
In the framework of the IMEGIN Large Program, we used the NIKA2 camera on the IRAM 30-m telescope to observe the edge-on galaxy NGC~891 at 1.15~mm and 2~mm and at 
a FWHM of 11.1$^{\prime\prime}$ and 17.6$^{\prime\prime}$, respectively.
Multiwavelength data enriched with the new NIKA2 observations fitted by the \texttt{HerBIE} SED code (coupled with the \texttt{THEMIS} dust model) were used to constrain the physical properties of the ISM. 
Emission originating from the diffuse dust disk is detected at all wavelengths from mid-IR to mm, while mid-IR observations reveal warm dust emission from compact \ion{H}{II} regions. 
Indications of mm excess emission have also been found in the outer parts of the galactic disk.
Furthermore, our SED fitting analysis constrained the mass fraction of the small (<~15~\AA) dust grains.
We found that small grains constitute 9.5\% of the total dust mass in the galactic plane, but 
this fraction increases up to $\sim20$\% at large distances (|z|~>~3~kpc) 
from the galactic plane.
}
\maketitle
\section{Introduction}
\label{sec:intro}

Dust modelling has often 
failed to explain the emission of galaxies at 
sub-millimeter (submm) to millimeter (mm) wavelengths. 
Excess emission has been detected in multiple studies (e.g., \cite{paradis+11,galametz+12,remyruyer+13,hermelo+16,galliano+18}) and its origin remains uncertain (see \cite{galliano+18} for a review). 
To explore this potential excess and its origin in various galactic environments, the nearly uncharted resolved mm emission is key.
The "Interpreting the Millimeter Emission of Galaxies with IRAM and NIKA2" (IMEGIN) 
Large Program (PI: S.~Madden)  is designed to decompose the interstellar medium (ISM) emission in this spectral domain. 
We are using the New IRAM Kid Arrays 2 (NIKA2) camera \cite{NIKA2-electronics,NIKA2-instrument,NIKA2-general,NIKA2-performance} on the \textit{Institut de Radio Astronomie Millim\'etrique} (IRAM) 30-m telescope to map 22 nearby galaxies (at distances smaller than 30~Mpc) at 1.15~mm and 2~mm.

A pilot study of IMEGIN has been the analysis of the edge-on galaxy NGC~891, at a distance of 9.6~Mpc \cite{bianchi+11}.
NGC~891 is considered a typical spiral galaxy with indications of more readily discernible 
structures (like, e.g., central bar, warped disk \cite{burillo+95}).
Because of its nearly perfect edge-on orientation, it is 
one of the most extensively observed nearby galaxies in a wide spectral range (e.g., \cite{guelin+93,seon+14,hodges-kluck+18}).
Previous studies have also 
investigated the dust emission (e.g., \cite{hughes+14,bocchio+16,yoon+21}) and the radio emission of the galaxy (e.g., \cite{mulcahy+18,schmidt+19}).
In this study we make use of this plethora of observations combined with the new NIKA2 observations in order to construct the spatially resolved Spectral Energy Distribution (SED) of the galaxy from mid-infrared (mid-IR) to radio wavelengths and put constraints on both the dust and radio emission of NGC~891.

\section{Methodology} \label{sec:metho}

\begin{figure}[!t]
    \centering
    \captionsetup{labelfont=bf}
    \begin{tabular}{@{}c@{}}
     \includegraphics[width=.51\textwidth]{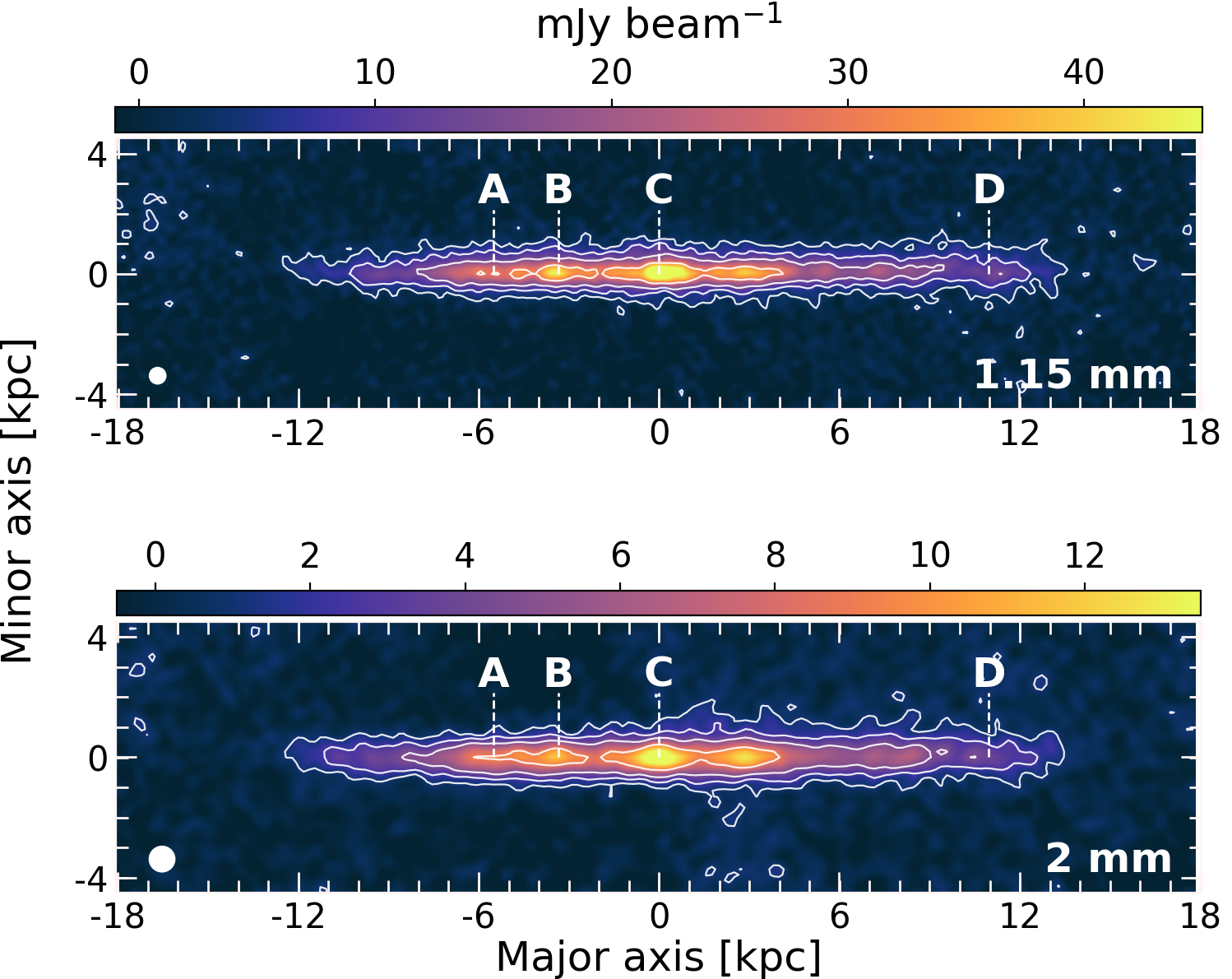} 
     \end{tabular}
    \begin{tabular}{@{}c@{}}
        \includegraphics[width=.44\textwidth]{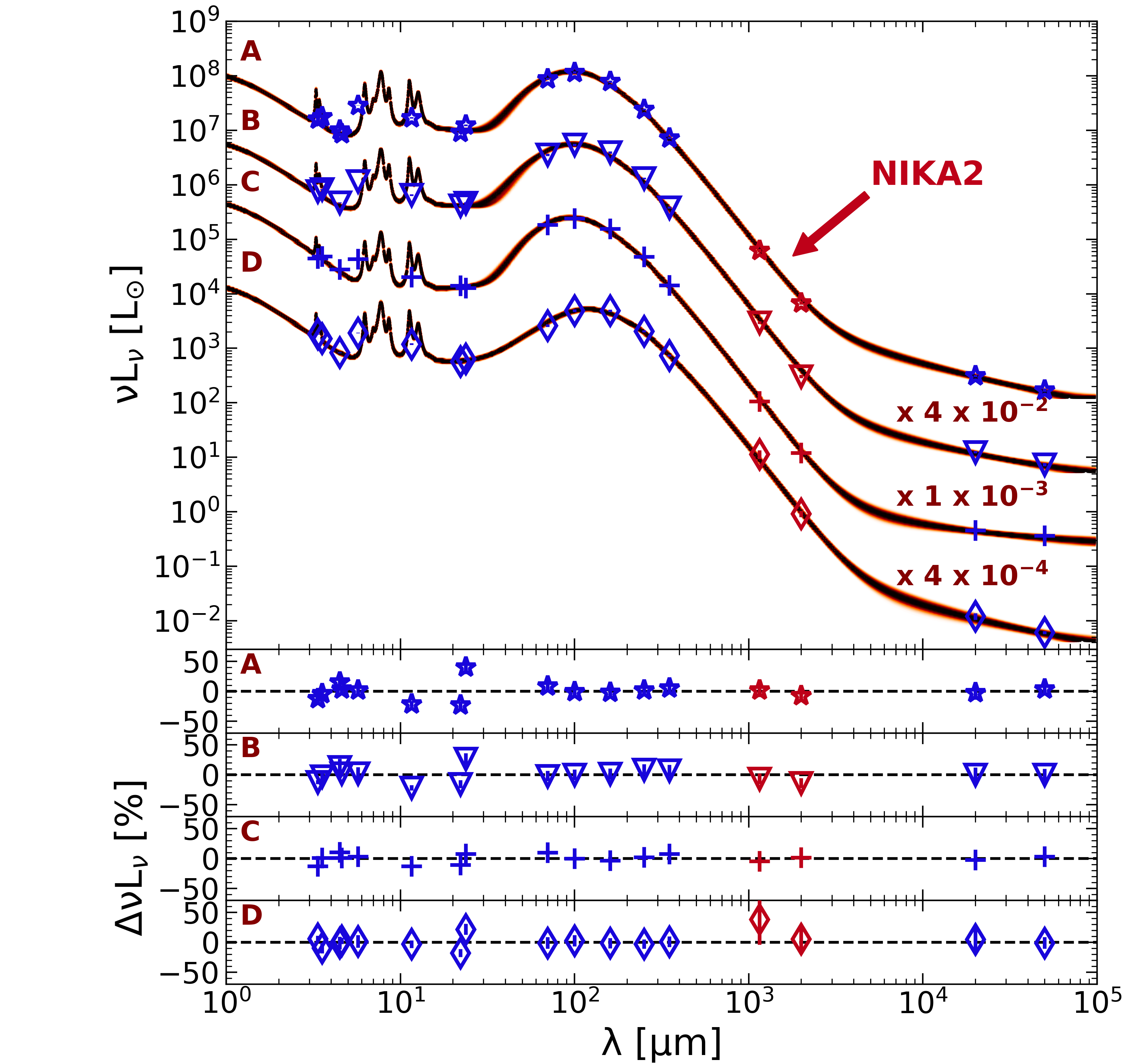} 
    \end{tabular}
    \caption{Left panel: NIKA2 observations of NGC~891 at 1.15~mm and 2~mm 
    with a beam FWHM of 11.1$^{\prime\prime}$ and 17.6$^{\prime\prime}$ respectively (beam sizes are indicated 
    with white circles in the bottom left corner). Four positions centered at -5.5, -3.4, 0.0 and 11.0~kpc along the major axis are annotated with the letters A, B, C and D respectively. Right panel: SEDs of the galaxy in the annotated regions A to D with each region covering an area of 8$^{\prime\prime}\times8^{\prime\prime}$ and fitted with the \texttt{HerBIE} code.}
    \label{fig:obs_sed}
\end{figure}

The current analysis uses new NIKA2 observations of NGC~891 at 1.15~mm and 2~mm, at a full width at half maximum (FWHM) of 11.1$^{\prime\prime}$ and 17.6$^{\prime\prime}$ (or linear resolutions of $\sim0.5$~kpc and $\sim0.8$~kpc at a distance of 10~Mpc), respectively. 
The NIKA2 camera operates simultaneously at 1.15 and 2~mm and needed 7 hours of total telescope time to map NGC~891.
The observations were combined and calibrated using the \texttt{piic/gildas} software\footnote{\href{https://publicwiki.iram.es/PIIC/} {\label{piic} https://publicwiki.iram.es/PIIC/}} and the final version of the calibration database with the data associated files (DAFs) \cite{zylka2013,berta-zylka2022}.
The final maps are presented in the left panel of Fig.~\ref{fig:obs_sed} centered at RA$_{J2000} = 2^h22^m33^s$, DEC$_{J2000}=+42^{\circ}20^{\prime}53^{\prime\prime}$ and rotated counter-clockwise by 67.1$^\circ$.

In order to construct the SED of the galaxy, the NIKA2 observations were combined with observations at wavelengths ranging from 3.5~$\mu$m to 5~cm and at FWHM up to 25$^{\prime\prime}$.
We used maps obtained with the \textit{Spitzer Space Telescope} (SST), the \textit{Wide-field Infrared Survey Explorer} (WISE), \textit{Herschel}, the \textit{Arcminute Microkelvin Imager} (AMI) and the \textit{Very Large Array} (VLA).
The background emission has been subtracted from the maps using the Python Toolkit for \texttt{SKIRT} (\texttt{PTS}) \cite{verstocken+20} framework.
In addition, the NIKA2 map at 1.15~mm has also been corrected for possible contamination from the 
CO(2-1) line, in accordance with the method described in \cite{drabek+12}.
Furthermore, following the methodology of \cite{verstocken+20} we generated the error maps of the observations.
We used the CO(3-2) map obtained from the \textit{James Clerk Maxwell Telescope} (JCMT) \cite{hughes+14} and the NIKA2 transmission curves reported in \cite{NIKA2-performance}.
Finally, all maps were homogenized to the same resolution of 25$^{\prime\prime}$, re-binned to a common grid with a pixel size of 8$^{\prime\prime}$ and masked applying a 3$\sigma$ cutoff.
The final fitted area includes the galactic disk and the extended emission of the galaxy up to $\sim5$~kpc from the midplane.

Using the homogenized maps, we performed a pixel-by-pixel SED analysis fitting with the Hierarchical Bayesian Inference for dust Emission (\texttt{HerBIE}) code \cite{galliano18,galliano+21}.
\texttt{HerBIE} is coupled with the \texttt{THEMIS} dust model \cite{jones+13,jones+17}, which takes into account realistic optical properties of the dust grains since it is based on laboratory data.
The code uses a hierarchical Bayesian approach aiming at reducing the noise-induced correlations between the inferred parameters, to recover their true correlations.
We fitted the multiwavelength data including the maps of the atomic and molecular hydrogen mass in the prior distribution. 
To trace these parameters we used the \ion{H}{I} \cite{oosterloo+07} and the CO(3-2) emission lines respectively.
In the right panel of Fig.~\ref{fig:obs_sed} we present four SEDs of the galaxy at the selected regions of interest A to D, annotated in the left frame.
These SEDs will be further discussed in the following sections.

\section{Discussion} \label{sec:discus}

The mm emission of NGC~891 measured in the NIKA2 bands (Fig.~\ref{fig:obs_sed}) has revealed regions of enhanced cold dust emission towards the inner parts of the galactic disk.
It shows a primary peak at the bulge of the galaxy (region C) and two secondary maxima at about $\pm3$~kpc from the galactic center along the major axis (e.g., region B).
The disk extension at both mm wavelengths reaches galactocentric distances up to $\sim13$~kpc.

\begin{figure}[!t]
\centering
\captionsetup{labelfont=bf}
\begin{tabular}{@{}c@{}}
\subfloat{\includegraphics[width=0.47\linewidth]{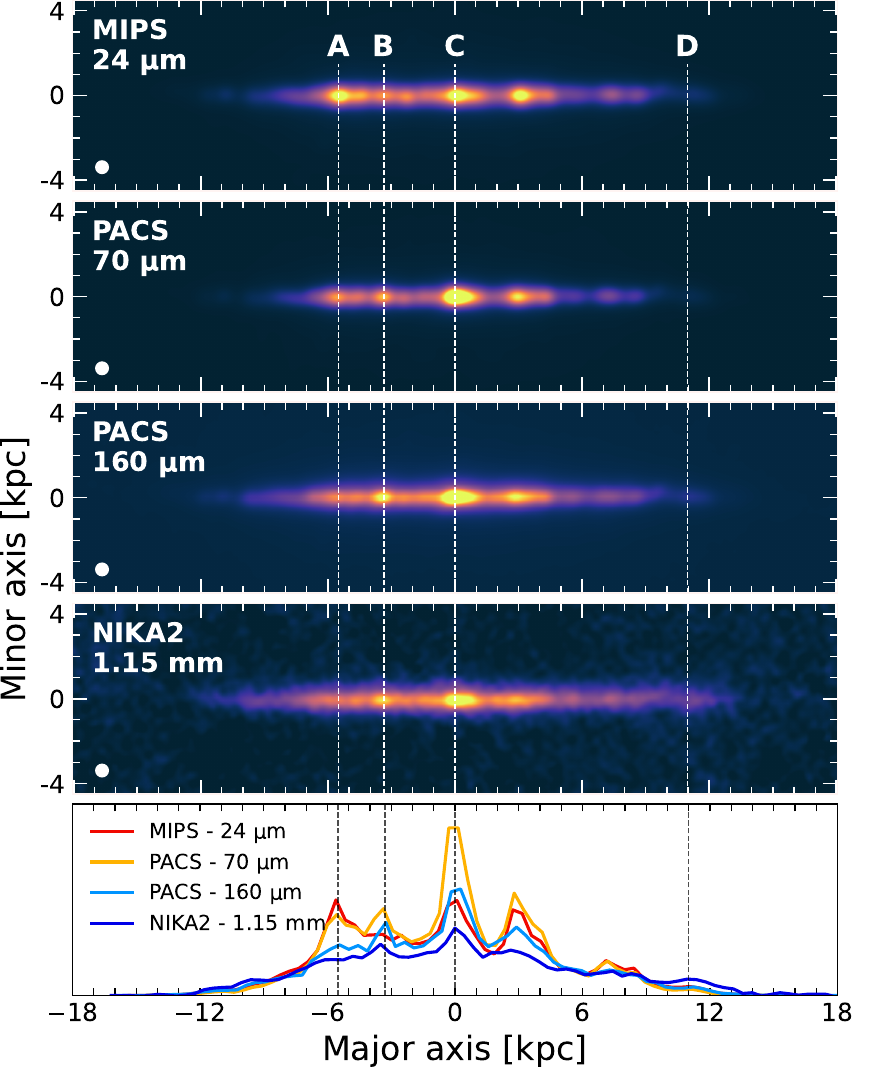}}\\ 
\end{tabular}\qquad 
\begin{tabular}{@{}c@{}}
\subfloat{\includegraphics[width=0.47\linewidth]{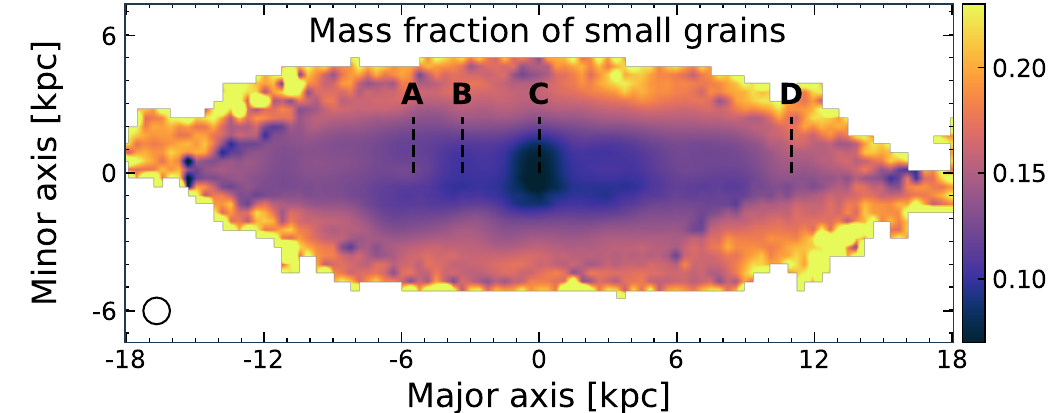}}\\
\subfloat{\includegraphics[width=0.47\linewidth]{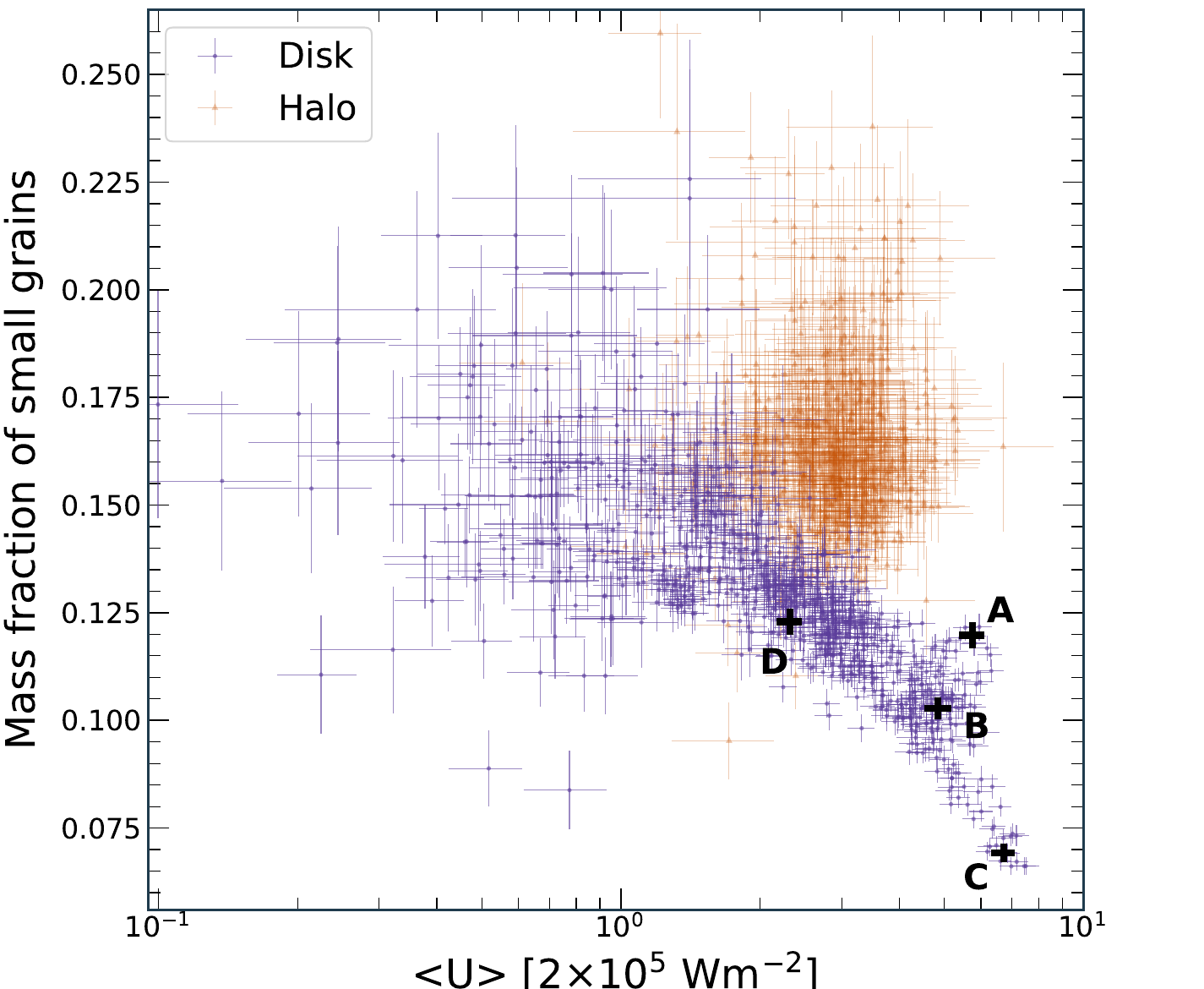}}\\ 
\end{tabular}
\caption{Left panel: maps of NGC~891 at 24~$\mu$m, 70~$\mu$m, 160~$\mu$m and 1.15~mm from top to the bottom, convolved to a FWHM of 12$^{\prime\prime}$, as tracers of different dust components. The radial profiles of the galactic emission at each wavelength, normalised at 6~kpc, are presented in the bottom panel. Right panel: the map in the top panel shows the mass fraction of small dust grains (<~15~\AA) in the galactic disk and halo, while in the bottom panel this parameter is plotted against the mean intensity of the ISRF (galactic plane pixels with purple crosses; halo pixels with orange crosses). Both parameters have been constrained through the SED fitting analysis.}
\label{fig:dust}
\end{figure}

\subsection{Dust environments along the galactic disk} \label{subsec:warm_cold}

Interstellar dust emits in a spectral range from mid-IR to mm wavelengths.
However, the intensity of dust emission in different galactic inner structures varies in a function of wavelength. 
NGC~891 maps at 24~$\mu$m, 70~$\mu$m, 160~$\mu$m, and 1.15~mm in the left frame of Fig.~\ref{fig:dust}, as well as their normalised intensity profile along the major axis in the bottom panel, show this clearly. 
The maps feature a bright inner disk with a central peak in the bulge of the galaxy (region C). 
Nevertheless, other disk substructures, such as regions A and B, show varying relative intensities at different wavelengths.
Region A is bright at 24~$\mu$m and becomes gradually dimmer at longer wavelengths.
On the contrary, region B is not prominent at 24~$\mu$m while becomes brighter at longer wavelengths.
The last annotated substructure, region D, is a special case with enhanced emission at 1.15~mm, but very faint emission at other wavelengths.

As warm dust is the major emission source at shorter wavelengths (e.g., 24~$\mu$m), while the emission at 160~$\mu$m and 1.15~mm trace on average cooler dust, we can discern different dust environments in each discussed region.
First, we detect emission from diffuse dust throughout the galactic disk.
Enhanced emission in the central bulge of the galaxy (region C) at all wavelengths comes from both warm and cold dust.
Cold dust dominates the emission in the secondary maxima of the galaxy (e.g., region B), where its spiral arms are projected to our line of sight.
And warm dust material is concentrated in region A, which may be considered a large \ion{H}{II} region. 
Finally, enhanced mm emission towards the outskirts of the galactic disk may indicate a very cold dust content.

\subsection{Decomposition into distinct dust grain populations}
\label{subsec:small_large}

Dust physical properties have been inferred through the SED analysis using the \texttt{THEMIS} dust model 
adapted to the \texttt{HerBIE} fitting code.
One of the main dust components of \texttt{THEMIS} is the aromatic-feature-emitting-grains (hereafter, small grains).
They are small, partially hydrogenated, 
amorphous carbons, a-C(:H), with radius smaller than 15~\AA, quite
similar to polycyclic aromatic hydrocarbons (PAHs). 
As a free parameter of the SED fitting, the mass fraction of small grains over the total dust mass is presented in the top right panel of Fig.~\ref{fig:dust}.
We can see that small grains are found in the galactic disk with mass fraction varying from $\sim6$\% in the center of the galaxy (region C) to $\sim15$\% in the regions of enhanced dust emission (e.g., regions A, B, D).
At high galactic latitudes the abundance of small grains increases up to $\sim20$\%.

In the bottom right panel we plotted the mass fraction of small grains against the mean intensity of the Interstellar Radiation Field (ISRF) for individual pixels of the galactic disk and the galactic halo (|z|~$\approx$~2~kpc).
The mass fraction of small grains in the galactic disk exhibits an anticorrelation with the ISRF intensity, as strong radiation fields destroy the small grains of a-C(:H).
Region A, already described as an \ion{H}{II} region with warm dust material, is also characterized by an enhanced abundance in small grains.
On the contrary, the galactic halo population occupies a different location in the plot with a steady intensity of ISRF and an enhanced small grains abundance with respect to the disk.
This excess in small grains may be caused by shuttering of larger carbon grains by shocks \citep[e.g.,][]{yamagishi+12}.

\section{Conclusions} \label{sec:conclu}

In this study we explored the properties of the ISM of NGC~891 performing resolved SED fitting and accounting for both dust and radio emission.
Comparing maps at different wavelengths we also distinguished different dust environments of the galactic disk.
An excess of mm emission of the galaxy has been reported towards the outer parts of the galactic disk.
Dust composition in small and large grains was investigated too.
We found that a fraction of $\sim$10\% of the total dust mass in the galactic disk and $\sim$20\% in the galactic halo 
comes from small dust grains and we explored its correlation with the intensity of the ISRF.
A further understanding of the ISM properties of galaxies requires similar studies in more IMEGIN galaxies \citep[see, e.g.,][]{katsioli+23}.

\section*{Ackowledgements}
\input{acknowledgements}

%
%
%

\end{document}

%% file: acknowledgements.tex
The research work was supported by the Hellenic Foundation for Research and Innovation (HFRI) under the 3rd Call for HFRI PhD Fellowships (Fellowship Number: 5357). We would like to thank the IRAM staff for their support during the observation campaigns. The NIKA2 dilution cryostat has been designed and built at the Institut N\'eel. In particular, we acknowledge the crucial contribution of the Cryogenics Group, and in particular Gregory Garde, Henri Rodenas, Jean-Paul Leggeri, Philippe Camus. The NIKA2 data were processed using the Pointing and Imaging In Continuum (PIIC) software, developed by Robert Zylka at the Institut de Radioastronomie Millim\'etrique (IRAM) and distributed by IRAM via the GILDAS pages. PIIC is the extension of the MOPSIC data reduction software to the case of NIKA2 data. This work has been partially funded by the Foundation Nanoscience Grenoble and the LabEx FOCUS ANR-11-LABX-0013. This work is supported by the French National Research Agency under the contracts "MKIDS", "NIKA" and ANR-15-CE31-0017 and in the framework of the "Investissements d’avenir” program (ANR-15-IDEX-02). This work has benefited from the support of the European Research Council Advanced Grant ORISTARS under the European Union's Seventh Framework Programme (Grant Agreement no. 291294). E. A. acknowledges funding from the French Programme d’investissements d’avenir through the Enigmass Labex. A. R. acknowledges financial support from the Italian Ministry of University and Research - Project Proposal CIR01$\_00010$. M. B., A. N., and S. C. M. acknowledge support from the Flemish Fund for Scientific Research (FWO-Vlaanderen, research project G0C4723N).